\documentclass[twocolumn,
showpacs,
preprintnumbers,
amsmath,
amssymb,
amsfonts,
final,
a4paper,
aps,
floatfix,
citeautoscript,
footinbib,
prl,
superscriptaddress]{revtex4}
\usepackage{times}
\usepackage{graphicx}
\usepackage[latin1]{inputenc}
\usepackage{color}
\usepackage{multirow}

\begin{document}
\title{Entanglement spectra of quantum Heisenberg ladders}
\author{Didier Poilblanc}
\affiliation{
Laboratoire de Physique Th\'eorique UMR5152, CNRS and Universit\'e de Toulouse, F-31062 France 
} 
\date{\today}
\begin{abstract}
Bipartite entanglement measures are fantastic tools to investigate quantum phases of correlated electrons. 
Here, I analyze the entanglement spectrum of {\it gapped} two-leg quantum Heisenberg ladders on a periodic ribbon
partitioned into two identical periodic chains. Comparison of various entanglement entropies proposed in the literature is given. 
The entanglement spectrum is shown to closely reflect the low-energy gapless spectrum of each individual edge, for any sign of the exchange coupling constants.
This extends the conjecture initially drawn for Fractional Quantum Hall systems to the field of quantum magnetism, stating a direct correspondence between
 the low-energy entanglement spectrum of a
partitioned system and the true spectrum of the {\sl virtual edges}.
A mapping of the reduced density matrix to a thermodynamic density matrix is also proposed via the introduction of an effective temperature.
\end{abstract}
\maketitle


{\it Introduction --} The recent application of quantum information concepts to several domains of condensed matter~\cite{RevModPhys} has proven to be extremely successful,
giving new type of physical insights on exotic quantum phases. 
Upon partitioning a many-body quantum system into two parts A and B, quantum entanglement can be characterized by the properties of 
the {\it groundstate} reduced density matrix of either one of the two parts, $\rho_A$ or $\rho_B$. 
For example, entanglement entropies such as the Von Neumann entropy $-{\rm Tr}\{\rho_A \ln\rho_A\}$ or the family of R\'enyi entropies
offer an extraordinary tool to 
identify a one-dimensional conformal invariant system~\cite{EntanglementEntropy} and provides e.g. a direct (numerical) calculation of its central charge~\cite{EE_numerics}.

Furthermore, the {\it entanglement spectrum} (ES) defined by the eigenvalues of a fictitious Hamiltonian $\cal H$, where $\rho_A$ is written as $\exp{(-\cal H)}$,
has been shown to provide much more complete information on the system.
In one dimension, underlying conformal field theory (CFT) leads to universal scalings of the ES~(Ref.~\onlinecite{OneDimension1})
and topological properties of the groundstate (GS) can be reflected by specific degeneracies~\cite{OneDimension2}. 
Choosing a partition corresponding to a very non-local real-space cut, the ES has also been used to define non-local order in {\it gapless} spin
chains~\cite{NonLocalOrder}.

\begin{figure}\begin{center}
  \includegraphics[width=0.9\columnwidth]{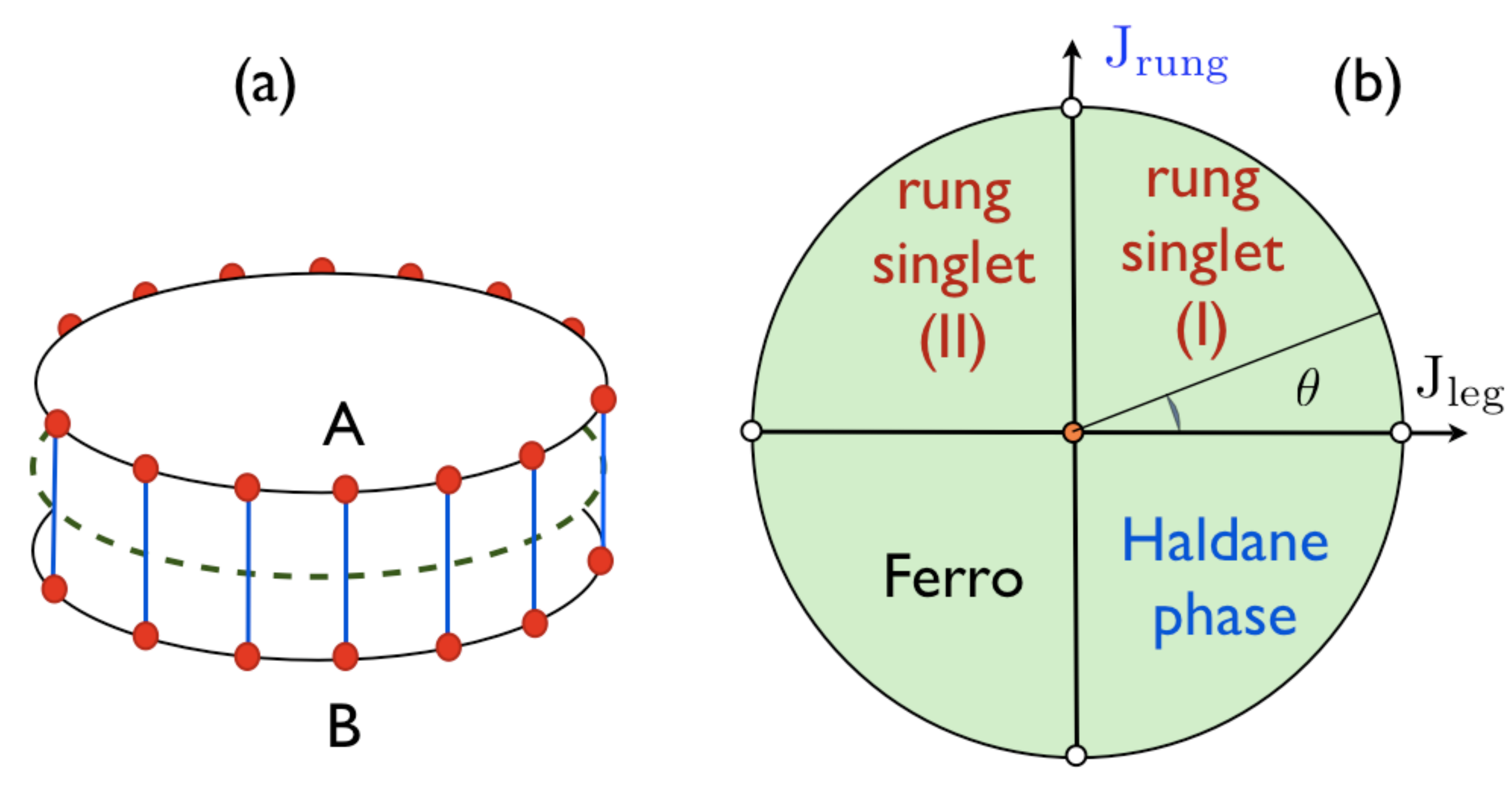}\end{center}  
  \caption{(Color online)
(a) Ribbon made of two coupled periodic Heisenberg chains (2-leg ladder). 
The partition into two identical A and B subsystems is made by cutting the rungs along the dashed line.
(b) Phase diagram of the two-leg ladder mapped onto a circle assuming $J_{\rm leg}=\cos\theta$ and $J_{\rm rung}=\sin\theta$.}
\label{Fig:ladder}
\end{figure}

Many-particle quantum entanglement is also a powerful tool to characterize topological features of two-dimensional GS~(Ref.~\onlinecite{EntanglementTopo2D})
as e.g. in dimer liquids on a cylinder geometry~\cite{DimerLiquidTopo}.
Also, bipartite ES have been shown to provide valuable informations on the edge states of fractional quantum Hall states 
on spherical~\cite{ES_NonAbelian} and torus geometries~\cite{ES_FQHedges} upon partition into two (identical) subsystems. 
Interestingly, the ES of  the incompressible GS of a generic Landau-level-projected Coulomb Hamiltonian arranges into a low-energy 
CFT spectrum, a fingerprint of topological order, separated by an `entanglement gap' from the high energy levels~\cite{ES_NonAbelian,EntanglementGap}.

Such advanced insightful analysis of the ES has not however been fully exploited in low dimensional quantum magnets.
In particular, the conjecture by Haldane of a precise correspondence between the entanglement spectrum and 
the true spectrum in reduced space, e.g. the spectrum of the subsystem A, is of very high interest and so far only supported by limited calculations
on quantum Hall systems.~\cite{ES_NonAbelian,ES_FQHedges} Low dimensional quantum magnets offer a completely different class
of many-body systems where new aspects of this correspondence can be investigated, giving further insights on this fascinating scenario.

{\it Model and System --} In this manuscript, I consider a 2-leg ladder made of two quantum Heisenberg  spin-1/2 chains coupled via a "rung" 
exchange coupling
$J_{\rm rung}$, as shown in Fig.~\ref{Fig:ladder}(a). 
Such a quantum magnetic ladder~\cite{ReviewLadders}
offers an attractive although still simple
system with three non-trivial phases, as shown in the phase diagram of Fig.~\ref{Fig:ladder}(b), depending on the signs of the leg (i.e. within the chains)
and rung Heisenberg exchange couplings, parametrized as $J_{\rm leg}=\cos{\theta}$ and $J_{\rm rung}=\sin{\theta}$ respectively. I shall not consider here the case 
when both couplings are ferromagnetic leading to a trivial fully polarized ferromagnet (lower-left quadrant). 
The physics of the other two phases (occupying the three remaining quadrants) can be easily understood starting from the strong rung coupling limit, i.e. when $|J_{\rm rung}|\gg J_{\rm leg}$.
When $J_{\rm leg}=0$ spin singlets or triplets form on the rungs depending whether the rung coupling is antiferromagnetic (AFM) or ferromagnetic (FM).
For an AFM rung coupling $J_{\rm rung}>0$, upon turning on a leg coupling of either sign
the product of rung singlets smoothly evolves into a (gapped) `rung singlet' phase.
For a FM rung coupling $J_{\rm rung}<0$ and a small AFM leg coupling $J_{\rm leg}>0$ the ladder system can be mapped onto an effective gapped spin-1 chain~\footnote{In the limit of vanishing $J_{\rm leg}$, 
the spin gap equals $\sim 0.41\times 2J_{\rm leg}$ and the spin correlation length is
$l_{\rm mag}\sim 6.01$. } yielding
an effective `Haldane phase' (Ref.~\onlinecite{Haldane,LadderTopo}). 
Remarkably, such gapped phases remain stable all the way to the weakly coupled chain regime.  
The rung coupling is therefore a `relevant' perturbation. For example, while the spectrum of the decoupled AFM chains system is the tensor product of
two gapless Conformal Field Theory (CFT) invariant low-energy spectra~\footnote{The Heisenberg spin-1/2 chain belongs to the SU(2)$_1$ Wess Zumino Witten universality class.} 
of central charge $c=1$, any finite $J_{\rm rung}$ opens a gap. 
Note also that the two rung singlet phases for AFM and FM leg couplings labeled as (I) and (II) in Fig.~\ref{Fig:ladder}(b) are smoothly connected to each other. 
Extension to {\sl frustrated} inter-chain couplings is considered in EPAPS~\cite{epaps}.

\begin{figure}
\begin{center}
  \includegraphics[width=0.9\columnwidth]{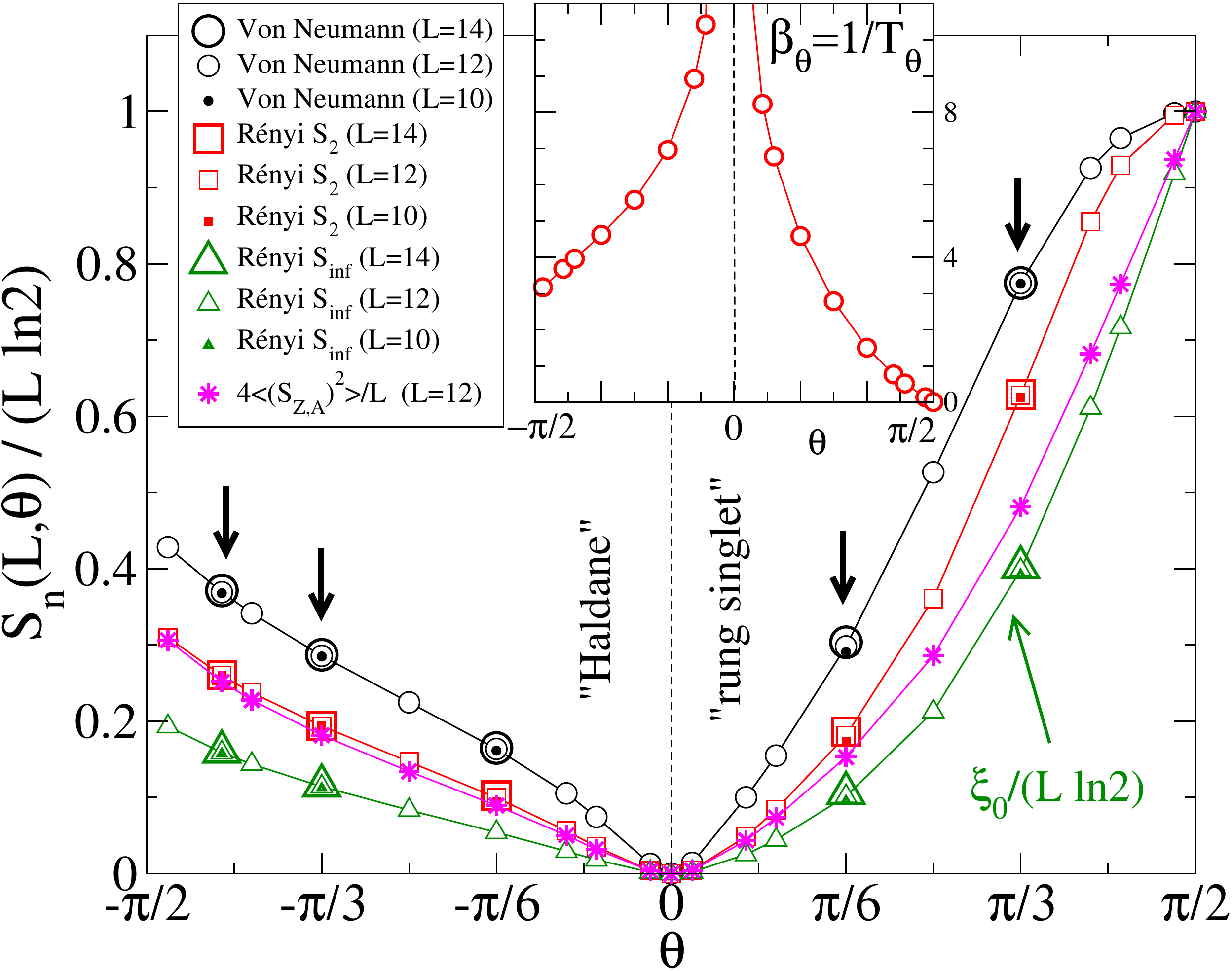}
\end{center}
\caption{(Color online) Various $S_n$ entanglement entropies ($n=1$ Von Neumann $S_{\rm VN}$ entropy, $n=2$ and $n=\infty$ R\'enyi entropies) 
computed on $2\times L$ ladders of length up to $L=14$, normalized by $L\ln 2$ and plotted versus the angle $\theta$. Note that $\xi_0=S_\infty\equiv S_{\rm inf}$. 
Only the two right quadrants of Fig.~\ref{Fig:ladder}(b) are considered. For comparison, the fluctuation of $S_{Z,A}$ (normalized by $L/4$) is also plotted (stars).
The corresponding ES are shown in Fig.~\ref{Fig:ES} for the values of $\theta$ marked by arrows. Inset: effective inverse temperature $\beta_\theta$ (see text).}
\label{Fig:entropies}
\end{figure}

The finite size two-leg ladder of Fig.~\ref{Fig:ladder}(a) is topologically equivalent to a ribbon which can be partitioned into two halves A and B
preserving periodic boundary conditions. This offers a simple convenient setup to investigate the entanglement between the two chain subsystems as a function of 
their coupling $J_{\rm rung}$. 
I report below the entanglement entropies as well as entanglement spectra in the two considered gapped phases, computed numerically on $2\times 10$, $2\times 12$ and $2\times 14$ 
clusters.  It is shown that the ES reflects the underlying CFT scaling behavior of the isolated chains.
This is remarkable, in particular in the strong rung coupling limit where the two subsystems are strongly entangled producing a short spin correlation length.
Note that I am considering here a different setup than the one used by Kallin et al.~\cite{VNentropyLadder} to calculate entanglement entropies on N-leg Heisenberg ladders.
Indeed, in the latter case, the A subsystem was chosen to have a `two-dimensional scaling' with its linear size.

{\it Results --} Characterizing the entanglement between A and B requires the knowledge of the reduced density matrix $\rho_A$ of the A subsystem (i.e. the upper AFM chain).
After computing the GS by Lanczos exact diagonalisation on finite $2\times L$ periodic clusters, an explicit use of translation symmetry enables
to express $\rho_A$ in a block-diagonal form, where each block corresponds to an irreducible representation labelled by one of the (allowed) total momentum $K=2\pi\frac{p}{L}$,
$p=-L/2+1,\cdots ,L/2$. 
These blocks can then be diagonalised (separately) to compute the Von Neumann (VN) entropy, $S_{\rm VN}=-{\rm Tr}\{\rho_A \ln \rho_A\}$, or the family of R\'enyi 
entropies,~\cite{RenyiEntropy}
$S_n=\frac{1}{1-n}\ln {\rm Tr}\{(\rho_A)^n\}$, $n\ge 2$. Note that $S_{\rm VN}$ can be considered as $ \lim_{n\rightarrow 1} S_n\equiv S_1$.
Results for $S_{\rm VN}$ and $S_2$ in the Haldane and rung singlet phases (for $J_{\rm leg}>0$) are reported in Fig.~\ref{Fig:entropies}.
The single-copy entanglement~\cite{SingleCopyEntanglement} obtained by taking the limit $n\rightarrow\infty$ and given by $S_{\infty}=-\ln \lambda_{0}$,
where $\lambda_{0}$  is the largest eigenvalue of $\rho_A$, is also shown for comparison.~\footnote{Let me recall that $S_{\rm VN}\equiv S_1\ge S_2\ge\cdots\ge S_\infty$.} 
An inspection of the finite size scaling of the data reveals that the leading term of all entanglement entropies
is proportional to the size $L$ (corresponding to the length of the edge between
A and B) as expected from the area law.
The data are therefore normalized by $L\ln 2$ which is the maximum (exact) entanglement entropy obtained for the product of independent rung singlets ($\theta=\pi/2$).
The finite size corrections (details in EPAPS~\cite{epaps}) are found to be very small, almost not visible at this scale. As also expected, all $S_n$ vanish in the limit of decoupled chains, where
the GS becomes a simple product state.
Interestingly, the 
behaviors of $S_{\rm VN}$ and $S_2$ are fairly similar, showing the same linear (quadratic) behavior
with $J_{\rm leg}\sim\Delta\theta$ in the strong rung coupling limit $\theta\rightarrow -\pi/2$ ($\theta\rightarrow \pi/2$).  
It should be noticed that, in contrast to  $S_{\rm VN}$ and $S_2$, $S_\infty$ behaves linearly when $\theta\rightarrow \pi/2$, a behavior also seen
in the quantum fluctuation of the (z-component of the) total spin $S_{Z,A}$ of the A subsystem. ~\cite{Fluc_1D}

\begin{figure}
\begin{center}
  \includegraphics[width=0.9\columnwidth]{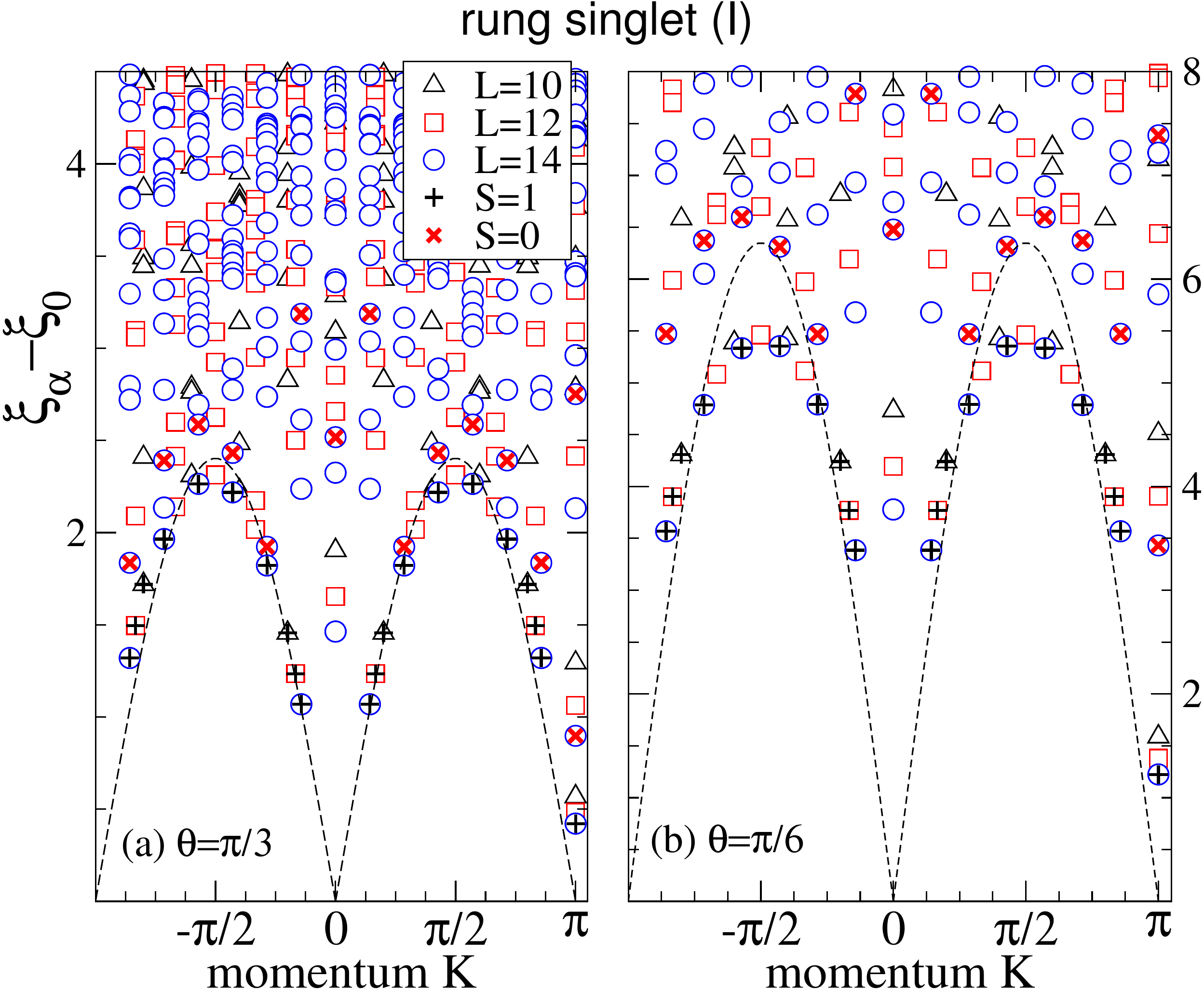}
  \includegraphics[width=0.9\columnwidth]{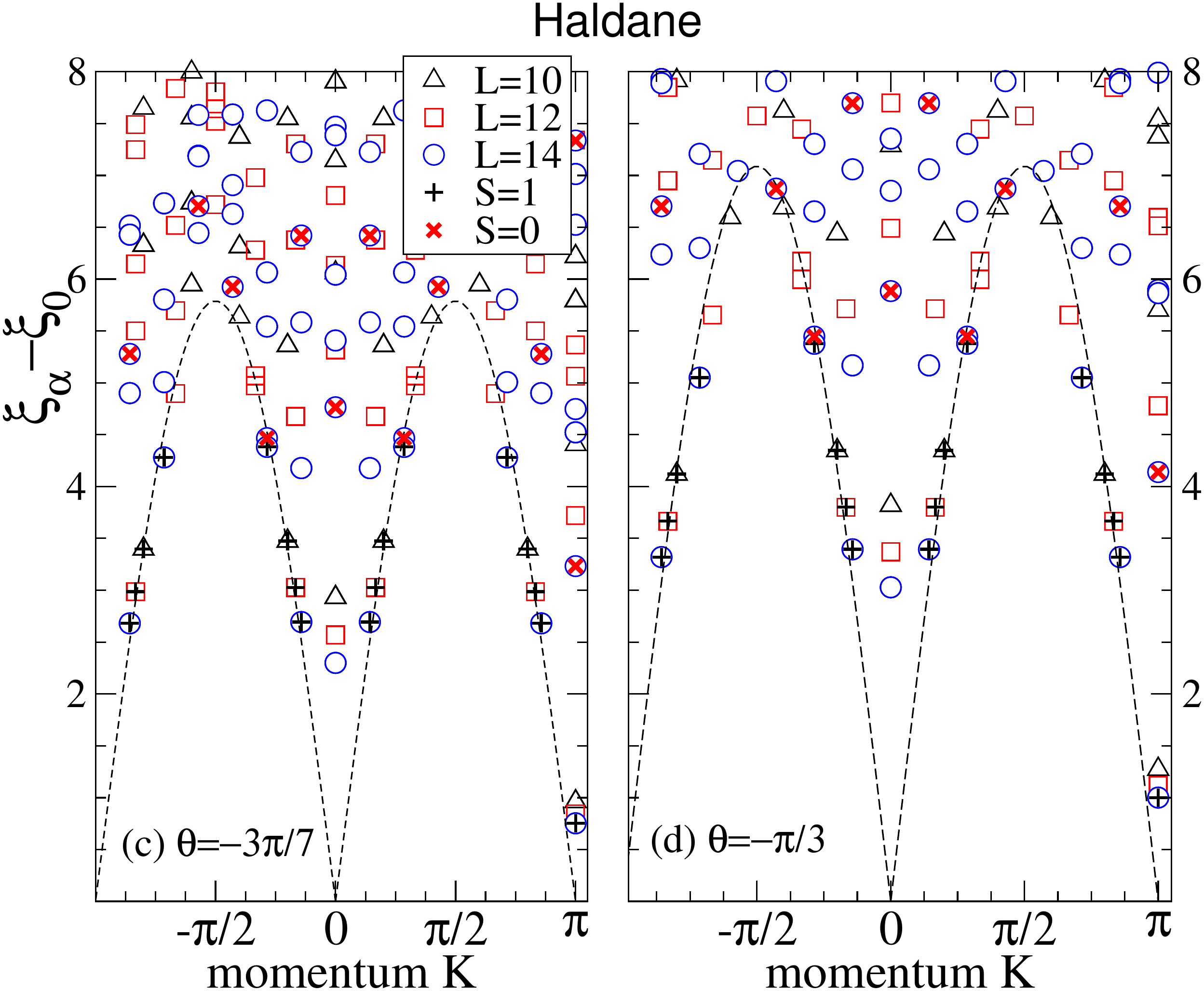}
\end{center}
\caption{(Color on line)
Entanglement {\it excitation} spectra versus total momenta $K$ in the chain direction for four different values of $\theta$ (shown by arrows in Fig.~\ref{Fig:entropies})
corresponding to the
rung singlet (I) phase (a,b) and the Haldane phase (c,d). All low-energy excitations computed on $2\times 10$, $2\times 12$ and $2\times 14$ ladders 
are shown by open (black) triangles, (red) squares and (blue) circles respectively. The lowest triplet eigenstates (for all $L$) are marked by (black) `$+$' symbols
and are fitted as $\Delta\xi=v|\sin{(K)}|$ by dashed lines.
The lowest singlet eigenstates for $L=14$ are also marked by (red) `$\times$' symbols.  
}
\label{Fig:ES}
\end{figure}

\begin{figure}
\begin{center}
  \includegraphics[width=0.9\columnwidth]{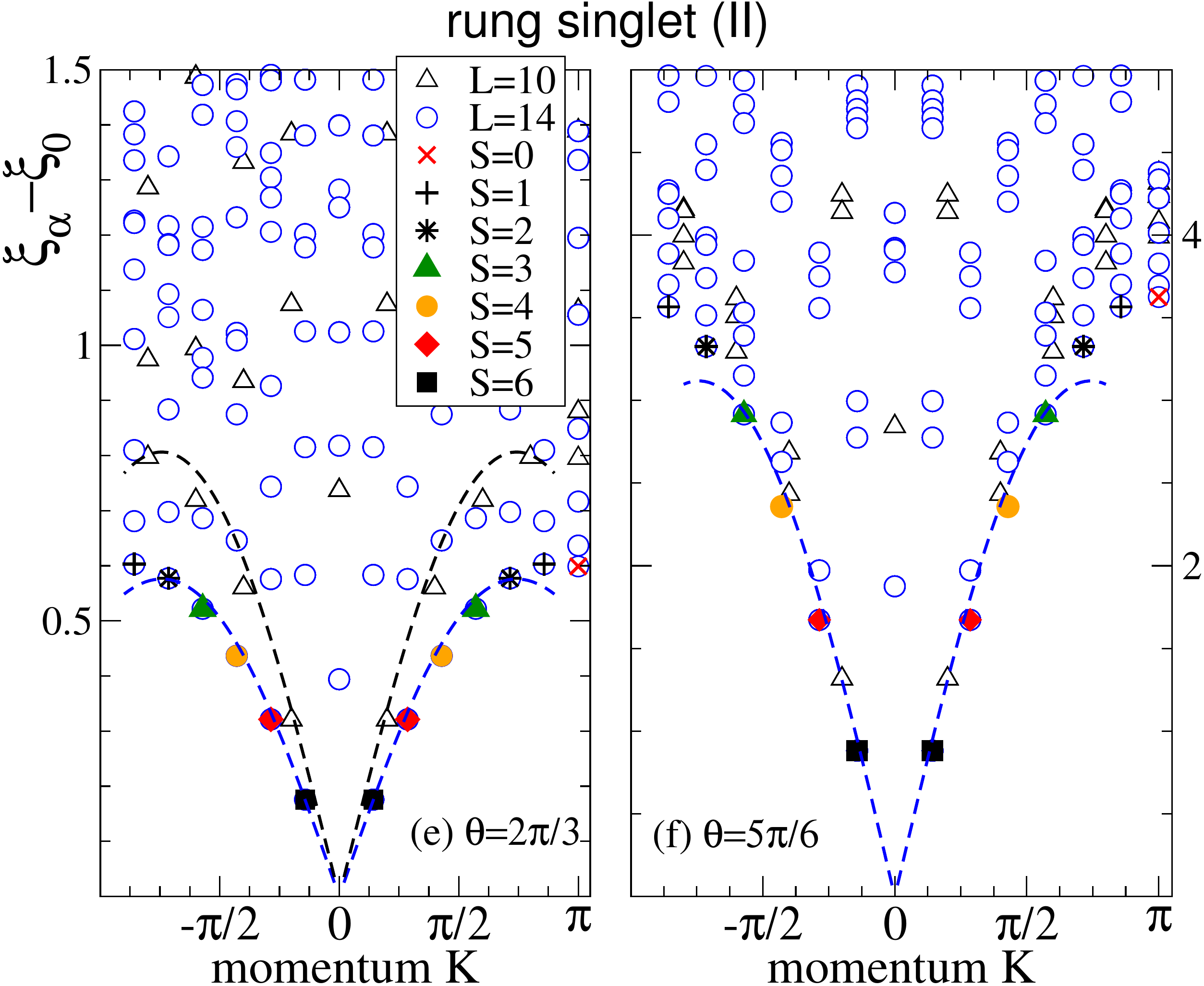}  
\end{center}
\caption{(Color on line)
(e,f) Same as Fig.~\ref{Fig:ES} for the rung singlet (II) phase (only $L=10$ and $L=14$ are shown). Here the GS is the saturated ferromagnet.
The total spin $S$ of the lowest eigenstates are indicated by different symbols (and colors)
and can be assigned to $m$-magnon bound states, $m=S_{\rm max}-S=L/2 - S$.
The lowest energy excitations for $L=14$ are fitted according to the formula for $E_{\rm min}(K)$ (see text). 
In (e),  the fit for $L=14$ rescaled by a factor $14/10$ (upper dotted line) gives also good agreement 
with the $L=10$ data.}
\label{Fig:ES2}
\end{figure}

I now move to the ES which is believed to contain much more information on the system. The ES is defined as the spectrum $\{\xi_\alpha\}$ of the 
hermician operator $\cal H$ given by the relation $\rho_A=\exp{(-\cal H)}$. The $\xi_\alpha$ can then be obtained from the weights $\lambda_\alpha$ of $\rho_A$ as
$\xi_\alpha=-\ln \lambda_\alpha$. 
Typical ES (measured from the GS energy $\xi_0$) plotted as a function of momentum $K$, for the three sizes,  are shown in Fig.~\ref{Fig:ES}, both in the rung singlet (a,b) and the 
Haldane (c,d) phases. Note that the total spin of the (A) subsystem is also a good quantum number which can be assigned to each level. It is remarkable that the low-energy excitations are 
spin-triplet that accurately resemble the des Cloiseaux--Pearson spectrum~\cite{HeisenbergChainTriplet} of the quantum Heisenberg chain (up to a multiplicative factor); in particular
two gapless modes at $K=0$ and $K=\pi$~\footnote{For $L=4p+2$, $K$ has to be shifted as  $K\rightarrow K-\pi$ to match the respective spectra with those for $L=4p$.} are clearly visible. 
The lowest singlet excitations close to $K=0$ and $K=\pi$ also form towers of states as predicted for the Heisenberg chain~\cite{HeisenbergChainSinglet}. 
This suggests strongly that the ES bears the {\sl same} low-energy CFT structure. In that case, we expect, in particular, the GS energy $\xi_0$ to scale as,
\begin{equation}
\xi_0/L= e_0 + d_1/L^2 + {\cal O}(1/L^3)\, .
\end{equation}
Such a behavior is indeed found for strong AFM rung couplings where $L\gg l_{\rm mag}$. 
Furthermore, the fit provides a number in good agreement with the CFT prediction $d_1=\pi c v/6$, where $v$ is the velocity of the triplet mode
and the central charge is set to 1, e.g. $d_1\simeq 1.31$ compared to $\pi c v/6\simeq 1.24$ for $\theta=\pi/3$.
However, for smaller rung couplings at which $L\sim l_{\rm mag}$ this scaling law is not satisfied (as expected)~\cite{epaps}.

I finish the investigation of the entanglement spectra by considering the last case of the rung singlet (II) phase realized for a {\it ferromagnetic} leg coupling ($J_{\rm leg}<0$) and
an AFM rung coupling $J_{\rm rung}>0$ (upper left quadrant of the phase diagram of Fig.~1(b)) and smoothly 
connected to the limit of decoupled rung singlets ($\theta=\pi/2$, $J_{\rm leg}=0$).
The results of the ES of $2\times 10$ and $2\times 14$ ladders are shown in Fig.~\ref{Fig:ES2}(e,f) for strong and weak rung couplings.
At low energies, the ES are shown to coincide (up to an overall rescaling factor) with the spectrum of the {\it ferromagnetic} quantum Heisenberg chain,
consisting of $m$-magnon bound states (or solitons)~\cite{Bethe} given by
$
E_m(K)= 2 J_{\rm eff} \sin^2{(K/2)}/m
$,
where $J_{\rm eff}$ is an effective chain coupling. 
On a finite cluster, such multi-magnon excitations are subject to the kinematic constraint $K\ge 2\pi m/L$, where $L$ is the ladder length. 
Therefore, the lower-bound energy `envelope'  behaves as,
$
E_{\rm min}(K)\sim \frac{4\pi}{L} J_{\rm eff} \sin^2{(K/2)} / K
$
up to small finite size corrections, as shown in Fig.~\ref{Fig:ES2} (note that $E_{\rm min}(K)\rightarrow 0$ for all $K$ in the thermodynamic limit). 

All these results on quantum ladders support the conjecture of a deep correspondence between the ES and the true spectrum of the (virtual) edges.

{\it Effective temperature  --}  Finally, I suggest that the reduced density matrix $\rho_A$ (in the GS) can be re-written as a thermodynamic density matrix by simply
introducing an effective, model-parameter dependent, temperature scale $T_\theta$ (focussing on the $J_{\rm leg}>0$ case). Indeed, comparison 
of Fig.~\ref{Fig:ES}(a) and Fig.~\ref{Fig:ES}(b) on one hand, and
of Fig.~\ref{Fig:ES}(c) and  Fig.~\ref{Fig:ES}(d) on the other hand, reveals almost identical spectra up to an overall multiplicative scale. This implies that $\rho_A$ can be written as,
\begin{equation}
\rho_A=\frac{1}{z_\theta}\exp{(-\beta_\theta \hat h)}\, ,
\label{Eq:thermal}
\end{equation}
where $\hat h$ is a parameter-free (extensive) Hamiltonian, $z_\theta=\lambda_0^{-1}$  and $\beta_\theta=T_\theta^{-1}$ is an effective inverse temperature to be adjusted. 
Since, as shown in Fig.~\ref{Fig:ES}, the spectrum of $\hat h$ has the same $c=1$ low-energy CFT structure
as the Heisenberg chain Hamiltonian
(up to a shift in GS energy) $\hat h$ can be `normalized'
by e.g. fixing the velocity $v$ of the triplet branch to be $v_{\rm Heis}=\pi/2$, the Heisenberg chain value. The effective inverse temperature $\beta_\theta$ 
is simply estimated from the actual slope of the gapless ($K=0$) mode of the corresponding ES and is reported in the inset of Fig.~\ref{Fig:entropies} as a function of $\theta$.
Apart from logarithmic corrections, the thermal (magnetic) length is $l_{\rm 1D}\sim T_\theta^{-1}$ (Ref.~\onlinecite{ThermalLength}) which, heuristically, can
be associated (up to a constant multiplicative factor of order 1) to the ladder  correlation length $l_{\rm mag}$. Therefore,  the behavior of $\beta_\theta$ in the inset
of Fig.~\ref{Fig:entropies} simply reflects the behavior of $l_{\rm mag}$ with $\theta$. In particular, in the strong AFM rung
coupling regime $J_{\rm rung}\gg J_{\rm leg}$, $\beta_\theta$ is linear in $\theta-\pi/2$, in agreement with the numerical estimation
of $l_{\rm mag}$, $l_{\rm mag}\propto J_{\rm leg}$.  More generally, within our normalization of $\hat h$,  $\beta_\theta\sim l_{\rm mag}/2 $.
Also, it is interesting here to use the exact equivalence between the entanglement VN entropy and the thermodynamic entropy of the (effective) finite-T subsystem.
In the regime of weakly coupled chains, using the expression of the thermodynamic entropy 
of the Heisenberg AFM chain when $T_\theta\ll 1$, one predicts $S_{\rm VN}/L \sim \pi T_{\theta}/(3v_{\rm Heis})$  (assuming again $c=1$)
which agrees (within less than $15\%$ difference) with the calculated 
VN entropy, giving further support that $\hat h$ belongs to the same universality class as the AFM Heisenberg chain (for $J_{\rm leg}>0$).

{\it Concluding remarks --} 
In this paper, I showed that the ES of the (ground state) reduced density matrix of a 2-leg quantum ladder possesses remarkable universal features
one can associate to its two single Heisenberg chain subsystems. 
This strongly supports a broader applicability (beyond quantum Hall systems) of the conjecture by Haldane establishing a deep
correspondence between the ground state ES of a many-body system made of two entangled constituents with the true spectra of the virtual {\sl edges}.
For example, although for AFM leg coupling the two ground states of the quantum ladder
at $\theta>0$ and $\theta<0$ belong to distinct topological sectors of the singlet spin Hilbert space~\cite{LadderTopo}  characterized by different `string orders',
it is remarkable that a similar $c=1$ CFT {\sl low-energy} ES is found and that Eq.~(\ref{Eq:thermal}) applies to both cases~\footnote{Subtle differences in the spectrum of $\hat h$ 
for $\theta>0$ and $\theta<0$ could probably be accounted for by including in $\hat h$  (small) deviations from a simple AFM Heisenberg chain Hamiltonian, not affecting the 
underlying CFT structure.}.
Similarly, although the two rung singlet phases (I) and (II) are smoothly connected, they exhibit completely different low-energy ES in straight connection to the different nature
of their edges. Lastly, I notice that the results of this paper also apply to the case of frustrated ladders~\cite{epaps}.


{\it Acknowledgements} -- I am indebted to S.~Capponi, N.~Laflorencie, G.~Misguich,  M.~Haque and P.~Pujol for interesting suggestions and/or comments. 
 I thank IDRIS (Orsay, France) for allocation of CPU time on the NEC supercomputer.

 
\end{document}


\title{Supplementary material to\\ Entanglement spectra of quantum Heisenberg ladders}
\author{Didier Poilblanc}
\affiliation{
Laboratoire de Physique Th\'eorique UMR5152, CNRS and Universit\'e de Toulouse, F-31062 France 
} 

\maketitle

This supplement of the main paper provides i) additional details and data regarding the scaling of the entanglement 
entropies and ii) extension of the calculations of ES to {\it frustrated} ladders. 

\section{Finite size scalings}

\begin{figure}[h,t,b]
\begin{center}
 \includegraphics[width=0.7\columnwidth,clip]{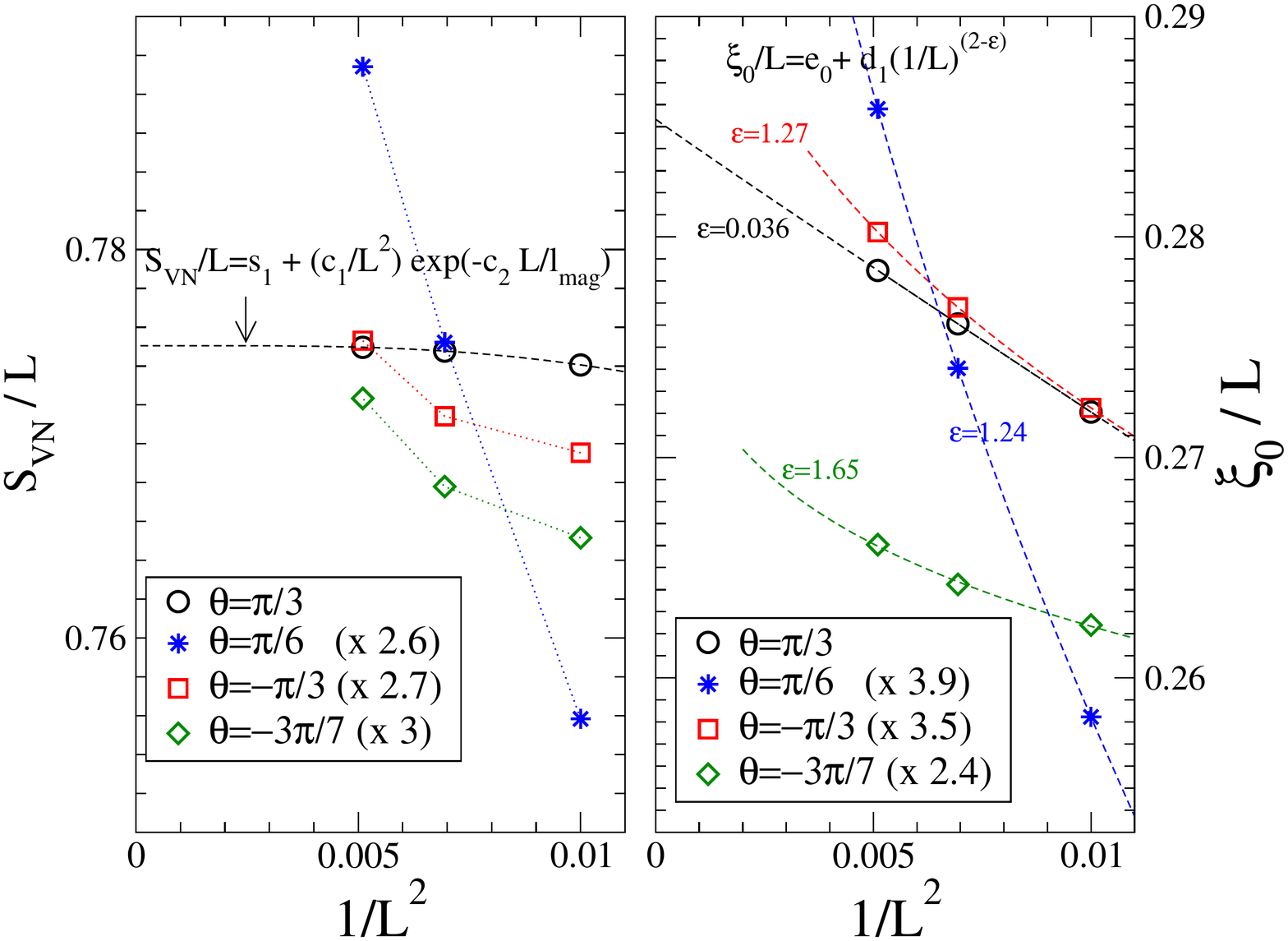}
\end{center}
\caption{
\label{Fig:scaling}
 Finite size scaling of the Von Neumann entanglement entropy (left) and of the GS energy $\xi_0$ of the ES spectrum (identical to the
single-copy entanglement  $S_\infty$) (right), both normalized by
system length $L$, as a function of $1/L^2$. Multiplicative factors have been used to bring all data in a single frame.
The values of $\theta$ are those indicated by the arrows in the Figure 2 of the main paper. 
The dashed ines correspond to an exponential fit in the left plot and to tentative power-law fits in the right plot. }
\end{figure}

As reported in the main paper, finite size corrections of the entanglement entropies remain small. 
The finite size scaling of the Von Neumann entropy and of the single-copy entanglement 
are provided in Fig.~\ref{Fig:scaling}. 
In the regime of strong AFM rung coupling the magnetic correlation length $l_{\rm mag}$ is very short and $L\gg  l_{\rm mag}$.
In this regime, the entanglement entropy scales with system size as,
\begin{equation}
S_{\rm VN} / L = s_1 + \frac{c_1}{L^p} \exp{(-c_2 L/l_{\rm mag})}     \, ,
\label{Eq:VN}
\end{equation}
where the value of $p$ is not crutial to the quality of the fit for $L$ varying from 10 to 14 (typically $p\sim 1$ or $\sim 2$).
However, finite size corrections at intermediate and small rung couplings do not show the exponential asymptotic behavior.  Very likely, this is the sign of a cross-over 
regime for which $L\sim  l_{\rm mag}$.  Another possibility is that the subleading corrections in $S_{\rm VN}$ vs. $L$ contain, {\it in this regime}, a constant term giving a $1/L$ contribution in Eq.~(\ref{Eq:VN}).  

\section{Frustrated interchain couplings}

\begin{figure}
\begin{center}
  \includegraphics[width=0.7\columnwidth,clip]{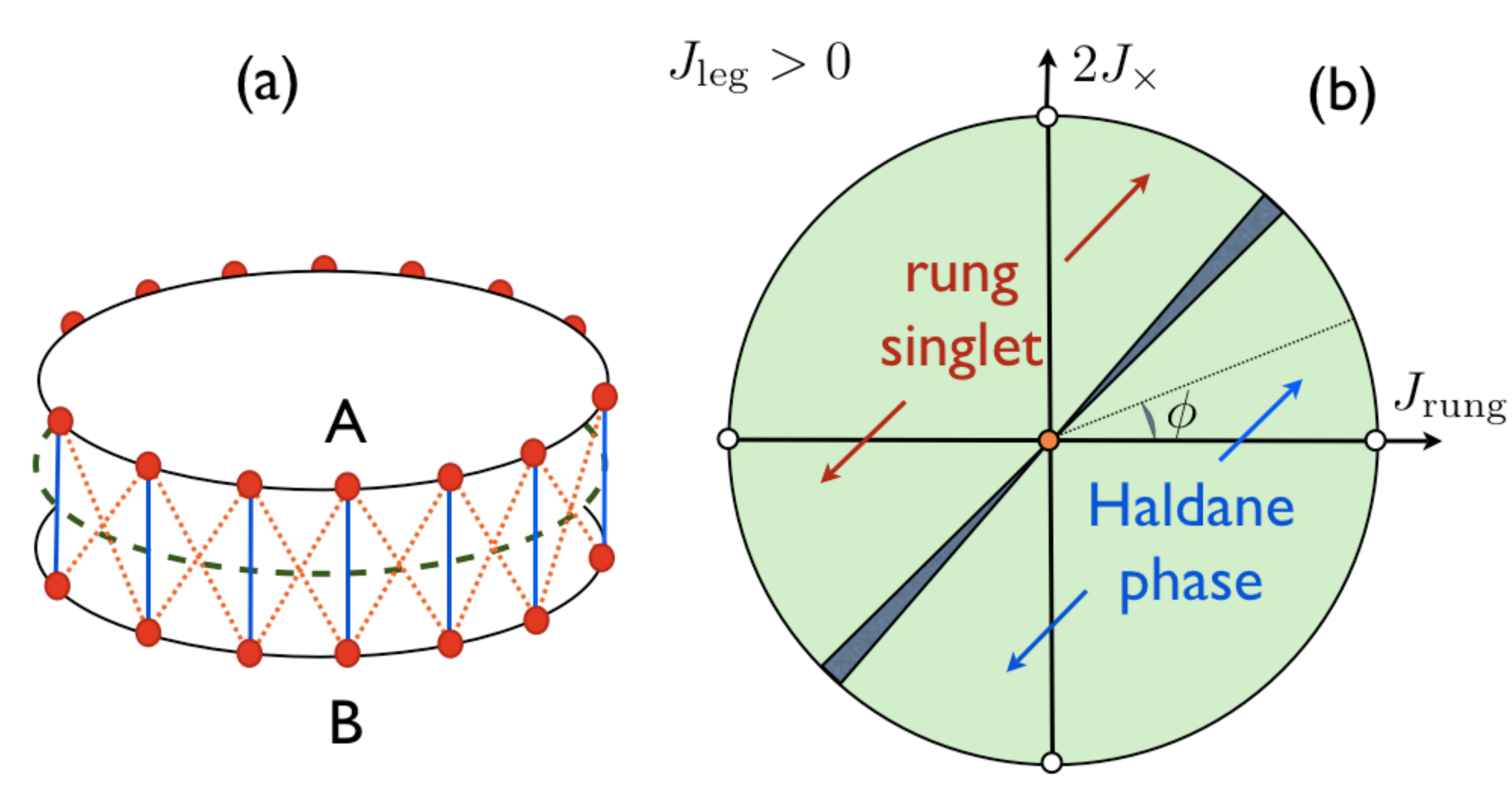}\end{center}  
  \caption{
(a) Ribbon made of two coupled periodic Heisenberg antiferromagnetic chains (2-leg ladder) coupled via vertical $J_{\rm rung}$ and diagonal 
$J_\times$ couplings. 
The partition into two identical A and B subsystems is made by cutting the rungs (and the diagonals) along the dashed line as in the main paper.
(b) Phase diagram of the two-leg ladder for a fixed value of the AFM leg coupling $J_{\rm leg}=\cos\theta>0$.
The parameter space $(J_{\rm rung},2J_\times)$ is mapped onto a circle of radius $\sin{\theta}$ i.e. 
$J_{\rm rung}=\sin{\theta}\cos{\phi}$ and $2J_\times=\sin{\theta}\sin{\phi}$.
Frustration occurs when these interchain couplings are both FM or AFM. The thin shaded areas schematically represent the (narrow) maximally frustrated regions
$2J_\times\sim J_{\rm rung}$ where a dimerized phase is expected. }
\label{Fig:ladder}
\end{figure}

Frustrated ladders as the one represented in Fig.~\ref{Fig:ladder}(a) have been investigated in the 
literature for both antiferromagnetic interchain couplings $J_{\rm rung}>0$ and $J_\times>0$~\cite{afm_frust},
and ferromagnetic interchain couplings $J_{\rm rung}<0$ and $J_\times<0$~\cite{fm_frust}. 
A schematic phase diagram is shown in Fig.~\ref{Fig:ladder}(b). In the two unfrustrated  quadrants 
(not studied here) we recover the same rung singlet and Haldane phases as in the main paper. 
A cross-over is expected in the frustrated regions corresponding to the two other quadrants
when $2J_\times\sim J_{\rm rung}$. Weak coupling Renormalisation Group treatments~\cite{afm_frust,fm_frust} 
predict the occurrence of a very narrow strip~\footnote{The intermediate phase is expected to be wider
for ferromagnetic interchain couplings due to the absence of marginal current-current interactions. See Ref.~\onlinecite{fm_frust}.} of a dimerized phase separated from the two limiting
rung singlet and Haldane phases by continuous transitions (i.e. quantum critical points). Numerical evidences
of this intermediate phase remain rather weak, except on the ferromagnetic side at large enough interchain
couplings. Note also that, at large enough {\it antiferromagnetic} interchain couplings, a direct first-order 
transition has been found, as in the classical limit. 

\begin{figure}
\begin{center}
  \includegraphics[width=0.7\columnwidth,clip]{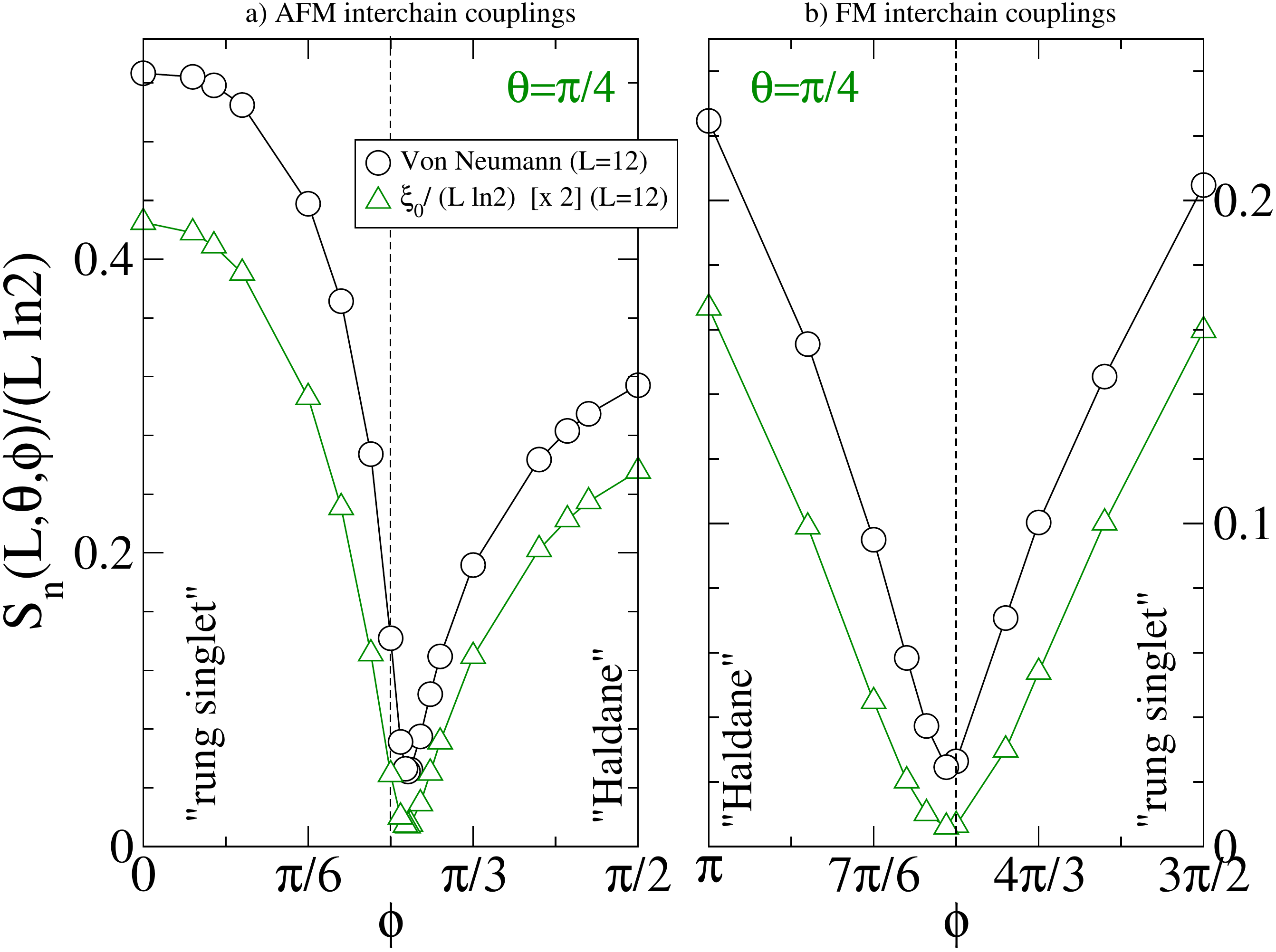}
\end{center}
\caption{
Von Neumann entanglement entropy $S_{\rm VN}$ entropy and  single-copy entanglement $S_{\infty}$
(i.e. R\'enyi entropy for $n=\infty$) computed on a $2\times 12$ ladder, normalized by $L\ln 2$ and plotted versus the angle $\phi$ (see Fig.~\ref{Fig:ladder}(b)) for fixed $\theta=\pi/4$. Note that $S_\infty=\xi_0$ (see main paper). 
Only the two opposite quadrants of Fig.~\ref{Fig:ladder}(b) exhibiting the narrow intermediate regions are shown. For convenience,
$S_\infty$ has been multiplied by a factor 2. 
Note the dip is more abrupt in the antiferromagnetic case. }
\label{Fig:entropies}
\end{figure}

The Von Neumann entropy as well as the $n=\infty$ R\'enyi entropy are shown in Fig.~\ref{Fig:entropies} 
in the two "frustrated" quadrants, for equal magnitudes of the exchange intra- and interchain couplings,  meaning
one fixes $\sqrt{J_{\rm rung}^2+J_\times^2}=J_{\rm leg}$ (i.e. $\theta=\pi/4$). As in the main paper, the entanglement 
entropies have been normalized with respect to
the ladder length $L=12$ (and divided by $\ln{2}$). The data clearly show a dramatic and rather sudden drop
of the entanglement entropies in the intermediate regions $2|J_\times|\sim|J_{\rm rung}|$.
This is reminiscent of the behavior obtained in the main paper when the two chains get decoupled as
$J_{\rm rung}\rightarrow 0$. The new remarkable result here is that the two chains {\it effectively} (almost) decouple
although the magnitudes of the intra- and interchain couplings remain equal 
(i.e. $\sqrt{J_{\rm rung}^2+J_\times^2}=J_{\rm leg}$) in the present example. On the other hand, 
the system size remains too small to reveal the presence of a {\sl finite} intermediate region. However, this region, if it exists,
should behave as {\it very weakly coupled} dimerized chains. Otherwise, it might correspond to a very exotic quantum critical point
(for antiferromagnetic interchain couplings). 

Finally, I discuss briefly the ES of the frustrated ladders. The results shown here are fully consistent with the "conjecture" discussed
at length in the main paper. In other words, the ES exhibit a generic low-energy CFT structure ($c=1$) similar to
the one of an isolated AFM Heisenberg chain. The ES hence enables to characterize the "edges" of the frustrated ladder. 
